# Life Between the City and the Village: Comparative Analysis of Service Access in Indian Urban Slums

Anand Sahasranaman[1,2,*] and Luís M. A. Bettencourt[3,4,#]

[1] Division of Mathematics and Computer Science, Krea University, Andhra Pradesh 517646, India.

[2] Centre for Complexity Science, Dept of Mathematics, Imperial College London, London SW72AZ, UK.

[3] Mansueto Institute for Urban Innovation, Dept of Ecology and Evolution, Dept of Sociology, University of Chicago, Chicago IL 60637, USA.

[4] Santa Fe Institute, Santa Fe NM 87501, USA

[*] Corresponding Author. Email: anand.sahasranaman@krea.edu.in

[#] Email: bettencourt@uchicago.edu

**September 2 2019**

**Abstract (192 words):**

The emergence of India as an urbanized nation is one of the most significant socioeconomic and political processes of the 21$^{st}$ century. An essential feature of India's urbanization has been the growth and persistence of informal settlements (slums) in its fast-developing cities. Whether living conditions in Indian urban slums constitute a path to human development or a poverty trap is therefore an issue of vital importance. Here, we characterize census data using the framework of urban scaling to systematically characterize the relative properties of Indian urban slums, focusing on attributes of neighborhoods such as access to basic services like water, sanitation, and electrical power. We find that slums in larger cities offer systematically higher levels of service access than those in smaller cities. Perhaps as expected, we also find consistent underperformance in service access in slums in comparison with non-slum neighborhoods in the same cities. However, urban slums, on average, offer greater access to services than neighborhoods in rural areas. This situation, which we quantify systematically, may help explain why Indian larger cities have remained attractive to rural populations in terms of living standards, beyond the need for an economic income premium.

# 1. Introduction

As the world becomes increasingly more urban with an expected 66% of the global population in living in cities by 2050, it is apparent that the bulk of future urbanization will be focused in Asia and Africa (UN, 2014). Three countries - India, China, and Nigeria - are expected to account for 37% of the projected growth in global urban population between 2014 and 2050, adding about 404 million, 292 million, and 212 million new urban residents, respectively (UN, 2014). One of the features of the urban expansion in many developing countries is the prevalence of slums. United Nations (UN) Habitat estimates that over 860 million people, including about one-third of the urban population in Asia and Africa, lived in urban slums as of 2012-13 (UN-Habitat, 2012).

While there is no globally consistent definition, the UN operationally defines slums as "communities characterized by insecure residential status, poor structural quality of housing, overcrowding, and inadequate access to safe water, sanitation, and other infrastructure" (UN-Habitat, 2003). Slums have been sometimes theorised as a transitory phenomenon in the life-cycle of rural-urban migration: people migrate into informal housing to gain an initial foothold in a city and over time, thanks to economic growth, eventually transition into more formal housing and better living conditions (Frankenhoff, 1967; Turner, 1969; Glaeser, 2011). This suggests that while slums may appear as a static reality for significant transitional periods of time, as urbanization levels increase and higher levels of development are reached, we should expect increasingly slum-free cities.

Of course, such general processual optimism may be unwarranted in new contexts, especially given the present worldwide scale and scope of slums, which require solutions in about one generation not several as has happened in today's high-income cities and nations. The present empirical reality in the developing world may be taken to suggest instead that slums appear to be a more dynamical phenomenon, which have not only existed, but grown over decades and have often housed multiple generations – many of whom have never escaped the poverty associated with growing up in these neighborhoods (Marx, Stoker, and Suri, 2013). Given the emergence and persistence of slums as part of the urbanization process in developing cities, the study of slums across the world has become an area of significant focus, especially for sociology and economics research (Heijnen et al., 2015; Fox, 2014; Kamndaya et al., 2014; Deshmukh, 2013; Szántó et al., 2012; Srivastava et al., 2012; Gupta, Arnols, and Lhungdim, 2009; Vaid et al., 2007; Owusu, Agyei-Mensah, and Lund, 2007; Imparato and Ruster, 2003; Karn and Harada, 2002). The robustness of slum evolution across global urban contexts combined with the expected growth of cities in developing countries over the next few decades renders it imperative that we develop a more systematic understanding of slum settlements.

Our present work seeks to contribute in this direction by identifying general quantitative characteristics of slums through the comparative analysis of data for Indian cities in light of the framework of urban scaling (Bettencourt, 2013; Bettencourt et al., 2007). Urban scaling analyses the covariation of urban indicators with population size to measure a set of agglomeration effects that are empirically found to be characteristic of cities across national contexts, from the United States and Europe to China, Brazil, South Africa, and India (Sahasranaman and Bettencourt, 2019a; Bettencourt and Lobo, 2018; Brelsford et al., 2017; Bettencourt, Lobo, and Strumsky, 2007; Bettencourt et al., 2007).

An urban indicator $Y_i(t, N_i)$ for city $i$, with population $N_i(t)$ at time $t$ is described as:

$$Y_i(t, N_i) = Y_0(t) N_i^\beta e^{\xi_i(t)}, \tag{1}$$

where $Y_0(t)$ signifies systemic change on the attribute being measured across all cities under consideration, $\beta$ is the scaling exponent or elasticity of $Y_i$ relative to population at given *t*, $\xi_i(t)$ represents idiosyncratic deviations from the scaling law (beyond the systemic change $Y_0$). Specifically, $\xi_i(t)$ are scale-independent deviations of individual cities from the scaling relation:

$$\xi_i(t) = \ln \frac{Y_i}{Y_0 N_i^\beta}. \tag{2}$$

The exponent, $\beta$, is empirically observed to assume characteristic values for different kinds of indicators at the functional city (metropolitan area) level: $\beta \simeq 7/6$ for socioeconomic attributes such as income, innovation, and crime; $\beta \simeq 5/6$ for public and network infrastructure such as road length and water network; and $\beta \simeq 1$ for individual or household level infrastructure like number of houses or household electricity connections (Bettencourt, 2013; Bettencourt et al., 2007). We seek to use the urban scaling framework to assess how properties of slums vary with city size, especially by measuring associated prefactors and exponents and comparing them to these reference cases in other contexts.

While the urban scaling framework can be applied to explore slum settlements across the world, we focus here on informal settlements in Indian cities. Given the expectation that India will be a predominantly urban country by 2050 (UN, 2017) housing 14% of the global urban population (Swerts, Pumain, and Denis, 2014) in conjunction with the historical reality of poor rural migrants concentrating in slums as their initial port of call in Indian cities (Gupta et al., 1992; Majumdar, 1978), scientific characterizations of slum settlements are imperative - and now, for the first time, possible due to the availability of data on slum-level basic services across the entire nation in the Census of India 2011.

There exists a rich legacy of surveys, ethnographies and micro-studies focused on slums in individual Indian cities, exploring specific aspects of these neighborhoods such as access to basic services, health, and living conditions (Heijnen et al., 2015; Deshmukh, 2013; Srivastava et al., 2012; Gupta, Arnols, and Lhungdim, 2009; Vaid et al., 2007; Karn and Harada, 2002). A study covering eight large Indian cities found that slums, on average, have significantly poorer housing and sanitation, higher total fertility rates, and lower vaccination coverage than non-slum neighborhoods (Gupta, Arnols, and Lhungdim, 2009). Microstudies of slums in Mumbai find that a significant proportion of their populations live in cramped quarters (82%), access public water connections (65%), and use public toilets (84%) (Deshmukh, 2013), with resultant high incidence of water-borne diseases, such as diarrhoea (614 per 1,000), typhoid (68 per 1,000), and malaria (126 per 1,000) (Karn and Harada, 2002). Similar studies in Uttar Pradesh and Tamil Nadu find a high prevalence of stunting and malnutrition among slum children (Srivastava et al., 2012) and high infant mortality due to diarrhoea (Vaid et al., 2007), respectively. Slum surveys in Odisha find that over half of slum households access drinking water from a public source, and a similar proportion practice open defecation, yielding a correspondingly high rate of diarrhoea prevalence (79%) (Heijnen et al., 2015).

Overall, these studies paint a picture of poor access to basic public services such as drinking water and sanitation in slums all over India. However, differences in scale, scope, and methodology across these multiple studies makes meaningful general comparison across slum contexts impossible. Therefore, while these studies provide important insights, analysis, and nuance for specific contexts, what we still lack is a complementary rigorous, quantitative understanding of the system of urban slums in India. Building such understanding would require a systematic exploration of questions on the state of basic infrastructure in slums across India, variations in slum service provision across cities, and differences in access to such services between slums and the cities in which they are situated. Using the approach of urban scaling, we seek to quantify answers to these questions and aim to construct a general scientific framework for understanding urban slums and associated issues of sustainable human development in India more systematically.

## 2. Results

For the first time, the Census of India of 2011 provided granular evidence on basic services available in slums across all Indian cities (Census of India, 2011a). It defined a slum as a "compact area, of at least 300 in population or about 60-70 households, of poorly built, congested tenements, in unhygienic environments usually with inadequate infrastructure and lacking in proper sanitary and drinking water facilities" (Census of India, 2011b). At the aggregate level, the Census revealed that over one in six urban Indian households lived in slums and that over a third of this population was concentrated in 46 Urban Local Bodies (ULBs) with population of 1 million or more (Roy et al., 2014). Census slum data is provided at the level of the Urban Local Body (ULBs) for each state and covers basic service data, specifically (i) length of paved roads; (ii) number of private toilets (disaggregated into various types such as pit, pour/flush, service, and others); (iii) number of community toilets; (iv) number of public tap points or public hydrants for protected water supply; (v) number of domestic electricity connections; and (vi) number of non-domestic electricity connections (disaggregated into road lighting and others) (Census of India, 2011a). No detailed maps or locations of these neighborhoods are provided at this point alongside this information.

While the data available is at the level of ULBs, it is important to consider the fact that the framework of urban scaling is applicable to functional cities that cover "commute to work" areas, comprising together both places of work and residence, and that non-trivial agglomeration effects expressed as non-linear scaling parameters are not typically found for urban areas defined using other characteristics such as density or political boundaries (Cottineau et al., 2015; Arcaute et al., 2014). The conception of functional cities is approximated by Census of India in its definition of units of analysis termed Urban Agglomerations (UAs), which are categorised by spatial contiguity, non-rural occupations, access to infrastructure and amenities (Census of India, 2011c). Given this definition of Indian UAs, we aggregate ULB data to the level of UAs for our analysis, and consider here results for all UAs with population above 50,000 (available online at: https://bit.ly/2UwHH5M). This accounts for an urban population of over 171 million and a corresponding slum population of 33.6 million, meaning that close to one-fifth (19.6%) of the population in these cities, on average, resides in slums.

The first and most important general relationship analyses the scaling of slum population with city (UA) population size. We see that slum population shows average superlinear scaling ($\beta = 1.06$, 95% Confidence Interval ($CI$): $[0.92, 1.20]$) with city size, indicating that the slum populations increase, on average, more than proportionally with city population (Figure 1a). This means that when comparing slum populations in cities of different sizes, a city twice the size of another is expected to have, on average, 6% more of its population living in slums. When we plot the rank order of deviations from scaling law ($\xi_i$ in Figure 1b) we find, for example, that Mumbai ranks 7th, implying that its slum population is significantly higher than already superlinear scaling expectations. We also find that Hyderabad and Kolkata have higher slum populations than expected, and Chennai and Pune only marginally so. However, the cities of Surat, Bangalore, and Ahmedabad have lower slum populations than scaling predictions.

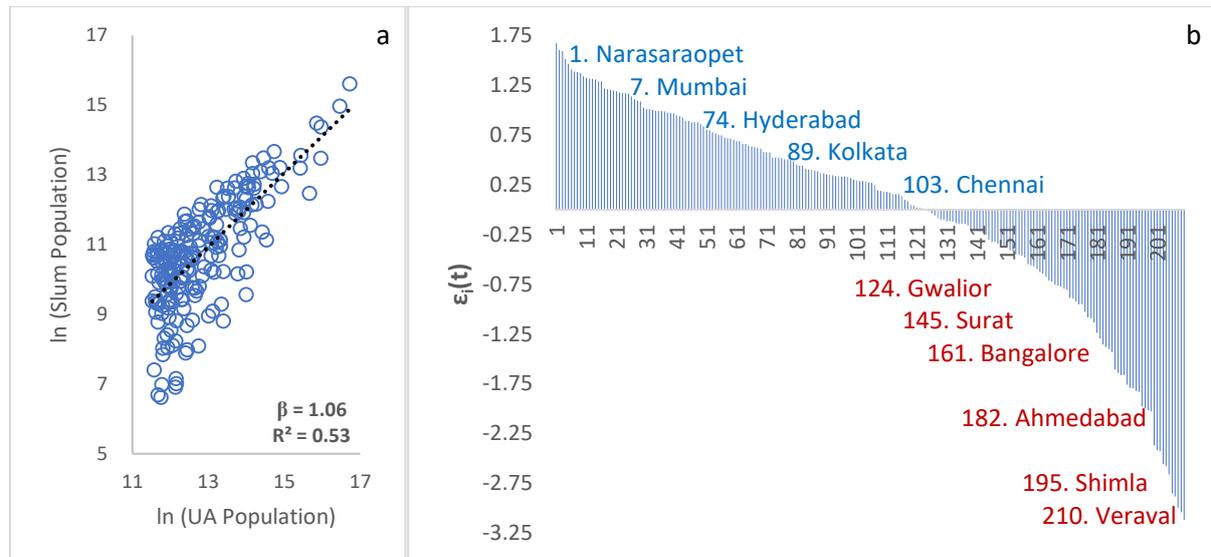

**Figure 1: Scaling and deviations of slum populations in UAs**: A: ln (slum population) versus ln (UA population) shows a superlinear relationship with β = 1.06 (95% CI: [0.92,1.20]). B: Rank order of UAs based on deviations from scaling relation, $\xi_i$. Mumbai, Kolkata, Hyderabad, Chennai and Pune have higher slum populations than superlinear scaling predicts, while Surat, Bangalore, and Ahmedabad have lower slum populations than expected.

Although the origin of this scaling effect of slum population with city size is complex and likely dependent on many factors, it is important to ask whether a greater relative slum population in larger cities is primarily the result of higher average GDP per capita (Sahasranaman and Bettencourt, 2019b), thus making larger cities greater economic magnets for migrants, or simply of the empirically observed scaling of basic services, which shows that larger cities, on average, offer better per capita basic service provision (Figure 2). Such a situation could be associated with a positive role of Indian cities in the transformation of (rural migrant) poverty into urban higher standards of living. But an alternative interpretation of the same facts, would be that observed conditions are not transient but instead more permanent (intergenerational) and to a large extent constitute poverty traps for large sectors of the urban population of India.

To better assess these issues, Figure 2 shows the scaling relationships for basic urban slum services with their total resident population. Because of the presence of zero values in the data for all basic services, we use logarithmic binning (logbinning) of data to measure these scaling relationships (Milojević, 2010). We pool data together from different UAs into 10 logbins, and for each of these logbins, compute the corresponding *log(average service count)*. The relationship between *log(slum population)* and *log(average service count)* is a measure of the percent change in per capita availability of a given service in slums with increasing slum population fraction. These relationships can then be compared to analogous scaling relations for the entire population.

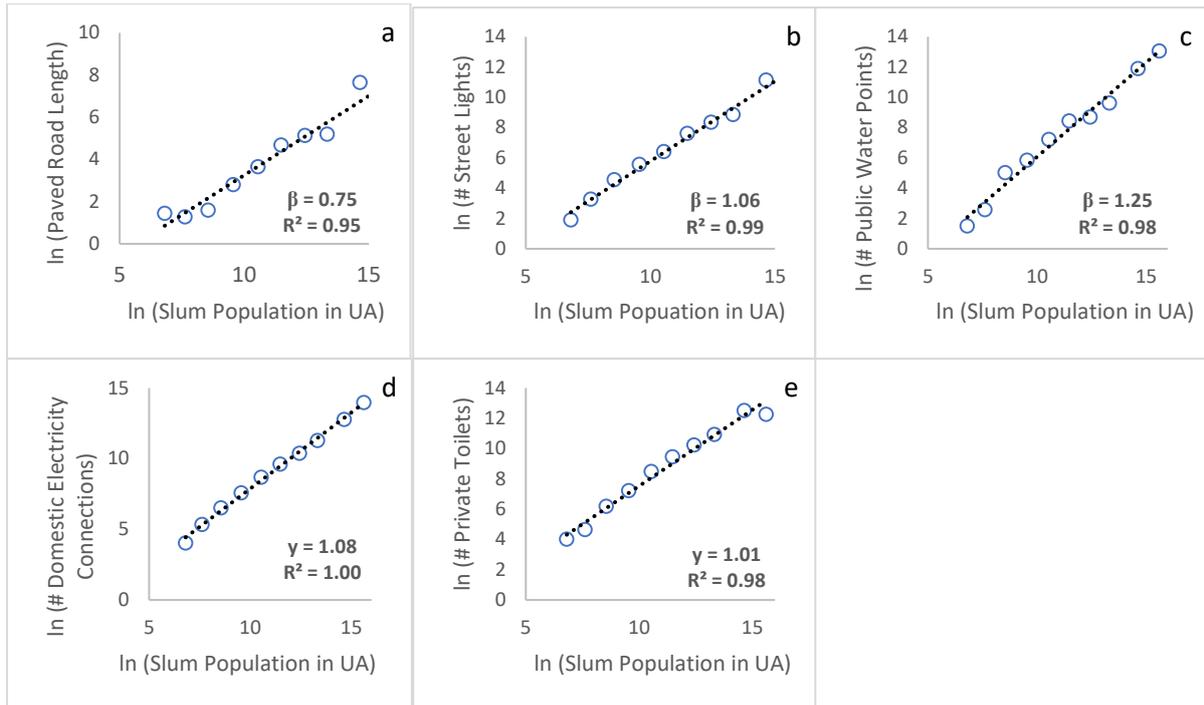

**Figure 2: Scaling of slum basic services with slum population in UAs:** A: ln (Length of paved slum roads) v ln (slum population in UA) shows sublinear scaling with exponent β = 0.75 (95% CI: [0.62,0.88]). B: ln (Number of street lights) v ln (slum population in UA) shows superlinear scaling with β = 1.06 (95% CI: [0.97,1.15]) revealing lower density of coverage by street lights in smaller cities. C: ln (Number of public water points) v ln (slum population in UA) shows superlinear scaling with exponent β = 1.25 (95% CI: [1.11,1.38]) meaning significantly higher access to public water points with increasing slum population. D: ln (Number of domestic electricity connections) v ln (slum population in UA) shows super linear scaling with β = 1.08 (95% CI: [1.02,1.13]). E: ln (Number of private toilets) v ln (slum population in UA) shows marginally superlinear scaling with β = 1.01 (95% CI: [0.89,1.13]). Overall, slums in larger cities offer improved access to basic services than those in smaller cities.

In terms of network infrastructure, we find that the length of paved slum roads shows a sublinear growth with slum population ($\beta = 0.75, 95\% \, CI: [0.62, 0.88]$), which is broadly in line with theoretical expectation for cities in general. Street lights scale slightly superlinearly ($\beta = 1.06, 95\% \, CI: [0.97, 1.15]$), reflecting an increasing lack of provision of this service in smaller cities (Figure 2a, 2b). We find significantly superlinear scaling of public water points with slum population ($\beta = 1.25, \, 95\% \, CI: [1.11, 1.38]$), indicating that a slum dweller in a city with a larger slum population has access, on average, to 25% more water supply points than one in a city that has half the slum population (Figure 2c). On private infrastructures, we find that the number of domestic electricity connections grows superlinearly with slum population, with an exponent $\beta = 1.08$ (95% $CI: [1.02, 1.13]$ (Figure 2d) and private toilets show marginal superliner behaviour, consistent with simple proportionality, with $\beta = 1.01$ (95% $CI: [0.89, 1.13]$ (Figure 2e).

These findings for basic household services in slums are different from what one typically finds for the same quantities in high income cities and nations, where linear scaling (Bettencourt, 2013) corresponds to basic service delivery that is essentially universal. In general, superlinear scaling of basic services is reflective of the fact that these basic services (water, sanitation, and electricity) are not yet universally available in India (Raghupathi, 2005) and that while access remains patchy across the spectrum of cities, slum dwellers in larger cities, with larger slum populations, have greater access. A systematic study of public service delivery in urban Indian slums clearly indicates that slums in larger cities offer consistently better measures of access to these basic facilities (Raghupathi, 2005). Specifically, the study classifies cities on the basis of population as metropolitan (pop. 1 million and above), Class I (pop. between 100,000 and 1 million), and Class II (pop. between 50,000 and 100,000), and finds, for instance, that per capita supply of water (in litres per capita per day) is 148 for metropolitan, 106 for Class I, and 69 for Class II cities; and the fraction of population dependent on low cost sanitation (pit latrines) unconnected to the sewerage network is 25% for metropolitan, 41% for Class I, and 55% for Class II cities (Raghupathi, 2005). Consequently, the rollout of basic service infrastructure in India follows on average down the urban hierarchy, both for slums and non-slums, a process also observed in other developing nations (Brelsford et al., 2017).

To investigate this point further, we now compare the scaling exponents and intercepts of slum infrastructure (scaling with slum populations) against those of overall UA infrastructure (scaling with total UA populations) to elicit the nature and extent of discrepancy in basic service provision in urban India (Table 1). While road length exhibits sublinear scaling for both slums and other overall city, there is a significant difference in exponents ($\beta_{city} = 0.96$, $\beta_{slum} = 0.75$) suggesting that roads scale significantly more sublinearly (less road per capita slum dweller) in slums. This could be an indication that despite the economies of scale in network infrastructure we would expect with increasing city size (Bettencourt, 2013), there is a systematic lack of provision of necessary road infrastructure in slums when compared to the cities they are in. This effect is also exacerbated by city size, not mitigated. Non-slum neighborhoods in cities are therefore significantly better provisioned with road accesses than slums, and as requisite roads are built in slums to address this deficit, we would expect this exponent $\beta_{slum}$ to rise to match the city at large. Again, this is typically a feature of informal settlements worldwide indicating deficits in other service delivery which tend to run along streets and accesses, as well as a lack of addresses and access to emergency services, including health and fire protection (Brelsford et al., 2018).

| Service outcome | Scaling of UA level service outcome with UA size | | Scaling of slum level service outcome with slum size | |
|---|---|---|---|---|
| | Intercept $Y_0(UA)$ | Exponent ($\beta_{city}$), with 95% CI | Intercept $Y_0(slum)$ | Exponent ($\beta_{slum}$), with 95% CI |
| Road length | 1.58 | 0.96 [0.92,1.01] | 0.01 | 0.75 [0.62,0.88] |
| Number of private toilets | 0.11 | 1.02 [0.96,1.07] | 0.08 | 1.01 [0.89,1.13] |
| Number of domestic electricity connections | 0.10 | 1.04 [1.01,1.07] | 0.05 | 1.08 [1.02,1.13] |

**Table 1:** *Comparison of scaling exponents and intercepts for UA-level and slum-level services.*

On private infrastructures, electricity connections exhibit similar superlinear scaling exponent ($\beta_{city} = 1.04$, $\beta_{slum} = 1.08$), meaning that domestic electricity connections scale similarly in cities and in their slums. It is apparent that both larger cities and larger slum populations in cities obtain proportionally greater access to domestic electricity connections. However, it is important to note that even while the scaling exponents are similar, there is substantive discrepancy in the intercepts: $Y_o(UA)$ is 1.8 times $Y_0(slum)$, meaning that the baseline level of provision of electricity in the UA is significantly higher (almost 2x) in cities than in slums.

Private toilets show similarity in scaling behaviour with slightly superlinear scaling for cities and slums ($\beta_{city} = 1.02, \beta_{slum} = 1.01$). This reflects the fact that per capita access to private toilets shows a slight overall increase with city/slum size. Again, while scaling exponents are similar, we compare the intercepts and find that $Y_o(UA)$ is 1.4 times $Y_0(slum)$, meaning that the baseline number of private toilets in the UA is again higher in cities than in slums (just as in the case of electricity connections). This finding is in agreement with Gupta, Arnols, and Lhungdim (2009), who studied eight Indian cities and found fewer and poorer sanitation facilities in slums versus non-slum neighborhoods, but also shows that the effect is indeed systematic, quantifiable, and applicable across all of urban India.

Prior work on scaling of various social and economic attributes in Indian cities has found strong regional discrepancies (Sahasranaman and Bettencourt 2019a, 2019b). Specifically, based on clustering analysis we concluded that cities in the South and West of the country tend to exhibit significantly different scaling behaviour than cities in the North, Centre and East. For slum characteristics, Figure 2 displays the scaling differences if we split the slum analysis by region – (i) Western and Southern (SW) cities (states of Gujarat, Mahasrashtra, Goa, Karnataka, Tamil Nadu, Andhra Pradesh and the Union Territory of Puducheri), and (ii) northern, central, and eastern (NCE) cities (Assam, Bihar, Chattisgarh, Haryana, Himachal Pradesh, Jammu and Kashmir, Jharkhand, Madhya Pradesh, Odisha, Punjab, Rajasthan, Uttar Pradesh, Uttarakhand, West Bengal and the Union Territory of Chandigarh). We find that availability of private toilets scales more superlinearly in SW cities than NCE cities (β =1.09 for SW cities, against 1.01 for NCE cities), but the baseline provision of toilets is higher in NCE cities ($Y_0$ of NCE cities is 2.4 times that of SW cities) (Figure 3a). The number of domestic electricity connections scales slightly more superlinearly in NCE cities (β=1.04 for SW cities, against 1.07 for NCE cities), but baseline for number of electricity connections is higher in SW cities ($Y_0$ of SW cities is 2.1 times that of NCE cities) (Figure 3b).

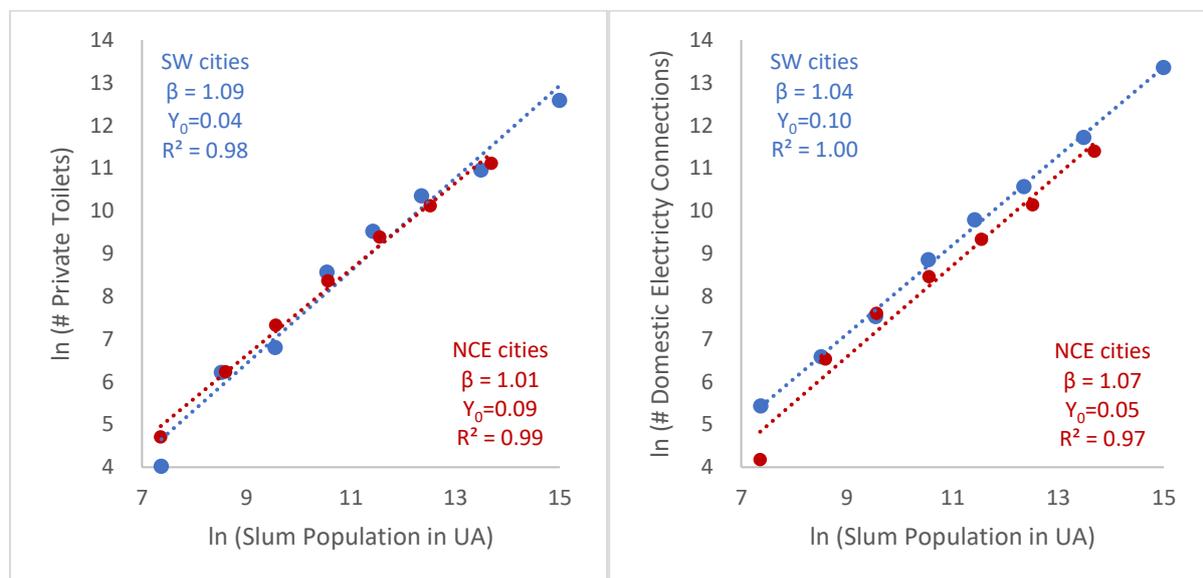

**Figure 3: Scaling slum basic services with slum population in UA by region**: A: ln (Number of private toilets) v ln (slum population in UA) shows marginally not significant superlinear scaling with β = 1.01 (95% CI: [0.91,1.10]) for NCE cities, and superlinear scaling with β = 1.09 (95% CI: [0.92,1.25]) for SW cities. The intercept for NCE cities is 0.09, while that for SW cities is 0.04. B: ln (Number of domestic electricity connections) v ln (slum population in UA) shows superlinear scaling with β = 1.07 (95% CI: [0.87,1.26]) for NCE cities, and superlinear scaling with β = 1.04 (95% CI: [1.00,1.08]) for SW cities. The intercept for NCE cities is 0.05, while that for SW cities is 0.10.

Finally, we provide another perspective on these relative deficits by comparing the levels of service provision in urban slums with rural areas. It has been argued that rural to urban migration may be driven by prospects of economic improvements, and that the urban poor are economically better off and happier than the rural poor (Glaeser, 2011). We attempt to assess these possibilities by characterizing how access levels to basic infrastructure in urban slums are different from access levels in rural India. Figure 4 clearly highlights that for both toilets (Figure 4a) and electricity (Figure 4b), urban slums, on average, provide better access than rural areas (red line). It is only in cities with the lowest (slum) populations that average service provision levels are lower than in rural areas. In choosing to migrate into larger urban areas therefore, rural households are also, on average, moving towards better access to basic services. However, it is also apparent that levels of service provision in urban slums are much lower than average levels of provision across all of urban India (blue line).

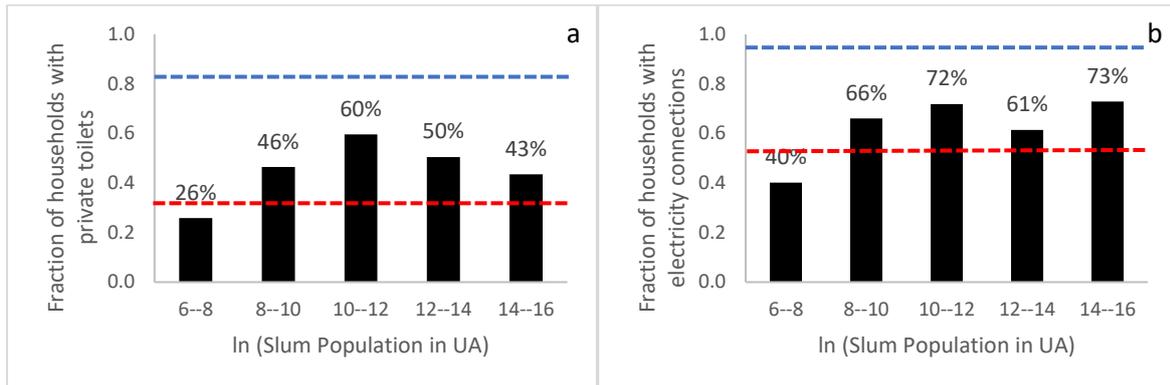

**Figure 4: Comparison of access to services between urban slums and rural areas:** A: Fraction of households with access to private toilets versus ln (Slum population in UA). Red Dashed Line: Average Fraction of rural households with private toilets = 31%; Blue Dashed Line: Average Fraction of urban households with private toilets = 81% (Census of India, 2011d). B: Fraction of households with access to domestic electricity connections v ln (Slum population in UA). Red Dashed Line: Average Fraction of rural households with electricity connections = 55%; Blue Dashed Line: Average Fraction of urban households with electricity connections = 93% (Census of India, 2011d). Only the UAs with lowest slum population appear to have lower service levels (both for toilets and electricity connections) than rural areas, but otherwise urban slums offer significantly better basic service delivery than rural areas.

Overall, urban scaling analysis of slum data covering all of India suggests that slums in larger cities are better equipped with basic services, such as electricity, water supply and sanitation than those in smaller cities. However, we find systematic discrepancies between slum and non-slum neighborhood in cities, especially in access to public roads, electrical connections, and private toilets, where slum areas are significantly under-provisioned. When comparing urban slums with rural areas, it is apparent that, on average, urban slums in larger cities offer far better access to basic services.

From a policy perspective, this suggests that there are two distinct aspects to understanding access to basic services in Indian urban slums. On the one hand, while slums in larger cities may be better served than those in smaller cities and rural areas, there appears to be the problem of significant lack of access to service across all scales. Additionally, within cities differences in service provision between slum and non-slum neighborhoods signal that urban segregation in Indian cities also occurs on the basis of place-based access to services as also observed in other nations (Brelsford et al., 2017), in addition to the more recognised aspects of socioeconomic status and caste (Singh and Vithayathil, 2012).

## 3. Conclusion

In this work we attempted to provide, to the best of our knowledge, the first systematic quantitative assessment of slums across all of urban India, through the comparative analysis of access to basic services in these neighborhoods relative to other relevant populations in the same cities and in rural areas. Slums have been studied by economists, sociologists and anthropologists exploring a wide variety of phenomena from access to services, to poverty, migration, and health. Here, we used the framework of urban scaling to produce a systematic, consistent comparative assessment of basic service delivery (water and sanitation, roads, and power) in Indian slums.

We find that slums in larger Indian cities have better access to public water, private toilets, and domestic electricity connections than those in smaller cities. This is in keeping with the revelation that slum population increases superlinearly with city size at present in India. We also find that basic services are significantly better provided in non-slum neighborhoods in the same cities. These systematic relationships mark a first step in creating a general scientific assessment of slums and neighborhoods in Indian cities, and establishing a benchmark for several dimensions of human development across this immense country. They suggest that basic service delivery in India is proceeding, on average, down the urban hierarchy, from larger cities to smaller ones, and that within each city it proceeds in a way that reflects place-based (dis)advantage, with slums showing relative deficits relative to non-slum populations. Thus, a process of sustainable urban development aimed at providing universal access to services in India, may stress the opportunities that arise with each types of inequity: i) between neighborhoods, and ii) across city size and levels of development. While solutions to the first benefit from economies of scale in infrastructure and service provision at the city scale, the second emphasizes the role of larger cities in the Indian urban hierarchy as places of institutional and technological innovation that can create and spread know-how to the rest of the nation.

It is important to point out that the Census of India has been criticized for likely undercounting slum populations and that the data analysed here is therefore almost certainly only an approximation to the real nature and extent of slum realities in India (Committee on Slum Statistics/Census, 2010). However, given the constraints on available consistent data on this issue, we believe the current analysis, with caveats, constitutes a relevant step in the development of a more systematic and deeper understanding of urban informality in India.

Given the expected fast rate of Indian urbanization over the next half century and the need for this process to become more sustainable, it will be critical to ensure the development of a general scientific understanding of the processes of human development in cities, including slum creation, their living conditions and the processes of human and urban development involved. We see our work exploring basic service delivery in slums through the prism of urban scaling as just a beginning in this direction, so that continued efforts to develop a science of cities that can account for the complex and dynamical realities of places like Indian cities will be essential to the creation of global sustainable urban futures.

**References:**

1. Arcaute E, Hatna E, Ferguson P, Youn H, Johansson A, Batty M. Constructing cities, deconstructing scaling laws. J R Soc Interface. 2014; 12: 20140745–20140745. doi:10.1098/rsif.2014.0745
2. Batty M. The Size, Scale, and Shape of Cities. Science. 2008; 319: 769–771. doi:10.1126/science.1151419

3. Bettencourt LMA. The Origins of Scaling in Cities. Science. 2013; 340: 1438–1441. doi:10.1126/science.1235823
4. Bettencourt LMA, Lobo J. Urban scaling in Europe. J R Soc Interface. 2016;13: 20160005. doi:10.1098/rsif.2016.0005
5. Bettencourt LMA, Lobo J, Helbing D, Kühnert C, West GB. Growth, innovation, scaling, and the pace of life in cities. Proc Natl Acad Sci. 2007; 104: 7301–7306. doi:10.1073/pnas.0610172104
6. Bettencourt LMA, Lobo J, Strumsky D. Invention in the city: Increasing returns to patenting as a scaling function of metropolitan size. Res Policy. 2007; 36: 107–120. doi:10.1016/j.respol.2006.09.026
7. Brelsford C, Lobo J, Hand J, Bettencourt LMA. Heterogeneity and scale of sustainable development in cities. Proc Natl Acad Sci. 2017; 201606033. doi: 10.1073/pnas.1606033114
8. Brelsford C, Martin T, Hand J, Bettencourt LMA. Toward cities without slums: Topology and the spatial evolution of neighborhoods. Sci. Adv. 2018; 4(8). doi: 10.1126/sciadv.aar4644
9. Census of India (2011a). http://www.censusindia.gov.in/2011census/dchb/DCHB.html. 2011.
10. Census of India (2011b). Primary census abstract for slum. Available at: www.censusindia.gov.in/2011-Documents/Slum-26-09-13.pdf. 2013.
11. Census of India (2011c). http://censusindia.gov.in/2011-prov-results/paper2/data_files/kerala/13-concept-34.pdf. 2011.
12. Census of India (2011d). Houses, Household Amenities and Assets Data 2001 – 2011: Visualizing Through Maps. Available at: www.censusindia.gov.in/2011-Common/NSDI/Houses_Household.pdf. 2011.
13. Committee on Slum Statistics/ Census. Report of the Committee on Slum Statistics/ Census. MoHUPA. 2010.
14. Cottineau C, Hatna E, Arcaute E, Batty M. Paradoxical Interpretations of Urban Scaling Laws. ArXiv150707878 Phys. 2015.
15. Deshmukh MS. Conditions of slum population of major sub0urban wards of Mumbai in Maharashtra. Voice of Res. 2013; 2(2).
16. Frankenhoff CA. Elements of an Economic Model for Slums in a Developing Economy. Econ Dev and Cultural Change. 1967; 16(1): 27–36.
17. Fox S. The Political Economy of Slums: Theory and Evidence from Sub-Saharan Africa. World Dev. 2014; 54: 191-203. doi: 10.1016/j.worlddev.2013.08.005
18. Glaeser E. Triumph of the City: How Our Greatest Invention Makes Us Richer, Smarter, Greener, Healthier, and Happier. 2011; New York, NY: Penguin Press.
19. Gomez-Lievano A, Youn H, Bettencourt LMA. The Statistics of Urban Scaling and Their Connection to Zipf's Law. PLoS ONE. 2012; 7: e40393. doi:10.1371/journal.pone.0040393
20. Gupta K, Arnold F, Lhungdim H. Health and Living Conditions in Eight Indian Cities. National Family Health Survey (NFHS-3), India. 2005-06. Mumbai: International Institute for Population Sciences; Calverton, Maryland, USA: ICF Macro
21. Gupta K, Nangia P, Fayazuddin M. Migrants in the Slums of Thane City. In Dynamics of Population and Family Welfare (eds.) Srinivasan K, Pathak KB. 1991, Bombay: Himalaya Publishing House.
22. Heijnen M, Routray P, Torondel B, Clasen T. Shared Sanitation versus Individual Household Latrines in Urban Slums: A Cross-Sectional Study in Orissa, India. Am. J. Trop. Med. Hyg. 2015; 93(2): 263–268. doi:10.4269/ajtmh.14-0812
23. Imparato I, Ruster J. Slum Upgrading and Participation: Lessons from Latin America. World Bank. 2003.

24. Kamndaya M, Thomas L, Vearey J, Sartorius B, Kazembe L. Material Deprivation Affects High Sexual Risk Behavior among Young People in Urban Slums, South Africa. J. Urb. Health. 2014; 91(3): 581-591. doi:10.1007/s11524-013-9856-1
25. Karn SK, Harada H. Field survey on water supply, sanitation and associated health impacts in urban poor communities – a case from Mumbai City, India. Water Sci. Tech. 2002; 46(11–12): 269-275.
26. Majumdar TK. The Urban Poor and Social Change: A Study of Squatter Settlements in Delhi. In The Indian City: Poverty, Ecology and Urban Development (eds.) D'Souza A. New Delhi: Manohar.
27. Marx B, Stoker T, Suri T. The Economics of Slums in the Developing World. J. Econ. Pers. 2013; 27(4): 187-210.
28. Milojević S. Power law distributions in information science: Making the case for logarithmic binning. J Am Soc Inf Sci Technol. 2010;61: 2417–2425. doi:10.1002/asi.21426
29. Owusu G, Agyei-Mensah S, Lund R. Slums of hope and slums of despair: Mobility and livelihoods in Nima, Accra. Norwegian J. Geog. 2008; 62(3): 180-190. doi:10.1080/00291950802335798
30. Raghupathi UP. Status of water supply, sanitation, and solid waste management in urban areas. New Delhi: National Institute of Urban Affairs; 2005.
31. Roy D, Lees MH, Palavalli B, Pfeffer K, Sloot MAP. The emergence of slums: A contemporary view on simulation models. Env. Modeling And Software. 2014; 59: 76-90.
32. Sahasranaman A, Bettencourt LMA. (2019a). Urban geography and scaling of contemporary Indian cities. J. R. Soc. Interface. 2019; 16. doi:10.1098/rsif.2018.0758
33. Sahasranaman A, Bettencourt LMA. (2019b). Economic geography and the scaling of urban and regional income in India. arXiv Preprint. 2019; arXiv:1902.09872.
34. Singh G, Vithayathil T. Spaces of discrimination. Econ. Pol. Wkly. 2012; 47(37).
35. Srivastava A, Mahmood SE, Srivastava PM, Shrotriya VP, Kumar B. Nutritional status of school-age children – A scenario of urban slums in India. Arch. Pub. Health. 2012; 70:8.
36. Swerts E, Pumain D, Denis E. The future of India's urbanization. Futures. 2014; 56: 43-52. doi:10.1016/j.futures.2013.10.008
37. Szántó GL, Letema SC, Tukahirwa JT, Mgana S, Oosterveer PJM, van Buuren JCL. Analyzing sanitation characteristics in the urban slums of East Africa. Water Pol. 2012; 14(4): 613-624. doi: 10.2166/wp.2012.093
38. Turner J. Uncontrolled Urban Settlement: Problems and Policies. In The City in Newly Developing Countries: Readings on Urbanism and Urbanization, (eds.) Breese GW. Englewood Cliffs, NJ: Prentice Hall; 1969. 507– 534.
39. UN, editor. World Urbanization Prospects: The 2014 Revision. United Nations, Department of Economic and Social Affairs, Population Division; 2015.
40. UN, editor. World Urbanization Prospects: The 2017 revision. United Nations, Department of Economic and Social Affairs, Population Division; 2017.
41. UN-Habitat. The Challenge of Slums: Global Report on Human Settlements 2003. London and Sterling, VA: Earthscan Publications; 2003.
42. UN-Habitat. State of the World's Cities 2012/2013: Prosperity of Cities. Nairobi: UNHSP; 2012.
43. Vaid A, Mammen A, Primrose B, Kang G. Infant Mortality in an Urban Slum. Ind. J. Pediatrics. 2007; 74.